\documentclass[12pt]{article}
\usepackage{amssymb}
\begin{document}



\def\a{\alpha}
\def\b{\beta}
\def\d{\delta}
\def\e{\epsilon}
\def\g{\gamma}
\def\h{\mathfrak{h}}
\def\k{\kappa}
\def\l{\lambda}
\def\o{\omega}
\def\p{\wp}
\def\r{\rho}
\def\t{\tau}
\def\s{\sigma}
\def\z{\zeta}
\def\x{\xi}
\def\V={{{\bf\rm{V}}}}
 \def\A{{\cal{A}}}
 \def\B{{\cal{B}}}
 \def\C{{\cal{C}}}
 \def\D{{\cal{D}}}
\def\G{\Gamma}
\def\K{{\cal{K}}}
\def\O{\Omega}
\def\R{\bar{R}}
\def\T{{\cal{T}}}
\def\L{\Lambda}
\def\f{E_{\tau,\eta}(sl_2)}
\def\E{E_{\tau,\eta}(sl_n)}
\def\Zb{\mathbb{Z}}
\def\Cb{\mathbb{C}}

\def\R{\overline{R}}

\def\beq{\begin{equation}}
\def\eeq{\end{equation}}
\def\bea{\begin{eqnarray}}
\def\eea{\end{eqnarray}}
\def\ba{\begin{array}}
\def\ea{\end{array}}
\def\no{\nonumber}
\def\le{\langle}
\def\re{\rangle}
\def\lt{\left}
\def\rt{\right}

\newtheorem{Theorem}{Theorem}
\newtheorem{Definition}{Definition}
\newtheorem{Proposition}{Proposition}
\newtheorem{Lemma}{Lemma}
\newtheorem{Corollary}{Corollary}
\newcommand{\proof}[1]{{\bf Proof. }
        #1\begin{flushright}$\Box$\end{flushright}}

\baselineskip=20pt

\newfont{\elevenmib}{cmmib10 scaled\magstep1}
\newcommand{\preprint}{
   \begin{flushleft}
   \end{flushleft}\vspace{-1.3cm}
   \begin{flushright}\normalsize
   \end{flushright}}
\newcommand{\Title}[1]{{\baselineskip=26pt
   \begin{center} \Large \bf #1 \\ \ \\ \end{center}}}
\newcommand{\Author}{\begin{center}
   \large \bf
Yupeng Wang${}^{a,b}$,~Wen-Li Yang${}^{c,d}$,~Junpeng
Cao${}^{a,b}$, ~Kangjie Shi${}^c$
 \end{center}}
\newcommand{\Address}{\begin{center}

     ${}^a$Beijing National Laboratory for Condensed Matter
          Physics, Institute of Physics, Chinese Academy of Sciences, Beijing
           100190, China\\
     ${}^b$Collaborative Innovation Center of Quantum Matter, Beijing,
     China\\
     ${}^c$Institute of Modern Physics, Northwest University,
     Xian 710069, China \\
     ${}^d$Beijing Center for Mathematics and Information Interdisciplinary Sciences, Beijing, 100048,  China
   \end{center}}
\newcommand{\Accepted}[1]{\begin{center}
   {\large \sf #1}\\ \vspace{1mm}{\small \sf Accepted for Publication}
   \end{center}}

\preprint
\thispagestyle{empty}
\bigskip\bigskip\bigskip

\Title{On the inhomogeneous T-Q relation for quantum integrable models}
\Author

\Address
\begin{abstract}
The off-diagonal Bethe Ansatz method \cite{wang} is used to revisit the periodic XXX Heisenberg spin-$\frac12$ chain. It is found that the spectrum of the transfer matrix can be characterized by an inhomogeneous T-Q relation, a natural but nontrivial extension of Baxter's T-Q relation \cite{baxter}.

\end{abstract}
\vspace{1cm}
\newpage
One of Baxter's important discoveries is the T-Q relation \cite{baxter}, which provides a convenient and universal
parametrization for eigenvalues of transfer matrices of most quantum integrable models. Nevertheless, the Q operator does not allow polynomial solutions for some integrable models such as the XYZ quantum spin chain with an odd number of sites and generic coupling constants, the chiral Potts model and the quantum spin chains with non-diagonal boundaries, despite the fact that the transfer matrix is a polynomial operator. It is
obvious that the T-Q parametrization for transfer matrix is not the unique one because
there are many ways to characterize a polynomial function, e.g.,
with its roots or with its coefficients. In a recent series of works (see ref.\cite{wang} and the references therein), a generalization of the T-Q relation with an extra inhomogeneous term, i.e., the inhomogeneous T-Q relation, was proposed and used in solving some integrable models without $U(1)$ symmetry. This generalization seems to be a universal solution of the Hirota type equations (recursive inversion identities) and can account for the boundary conditions self-consistently without losing a polynomial Q operator.

In this note, we show that the inhomogeneous T-Q relation can also characterize the spectrum of the ordinary integrable models that can be characterized by Baxter's T-Q relation and can be solved with the ordinary Bethe Ansatz methods.

Let us consider the periodic XXX spin-$\frac 12$ chain. The corresponding $R$-matrix reads
\begin{eqnarray}
R_{0,j}(u)=u+\eta P_{0,j}=u+\frac{1}2\eta(1+
\vec{\sigma}_0\cdot\vec{\sigma}_j), \label{20140317-r}
\end{eqnarray}
where $\eta$ is the crossing parameter (we put $\eta=1$ in this
case), $\vec{\sigma}_j=(\sigma_j^x, \sigma_j^y, \sigma_j^z)$ are the
Pauli matrices, and $P_{i,j}$ is the permutation operator possessing
the properties:
\begin{eqnarray}
&& P_{i,j}O_j=O_iP_{i,j},\quad  P_{i,j}^2={\rm id},\quad
tr_jP_{i,j}=tr_iP_{i,j}={\rm id},
\end{eqnarray}
for arbitrary operator $O$ defined in the corresponding tensor space. This $R$-matrix satisfies the Yang-Baxter equation
\begin{eqnarray}
R_{1,2}(u-v)R_{1,3}(u)R_{2,3}(v)=R_{2,3}(v)R_{1,3}(u)R_{1,2}(u-v).
\end{eqnarray}
It is easy to show that the $R$-matrix (\ref{20140317-r}) also satisfies the following relations:
\begin{eqnarray}
&&\mbox{ Initial condition}:\,R_{1,2}(0)= P_{1,2},\label{3Int-R}\\
&&\mbox{ Unitary relation}:\,R_{1,2}(u)R_{2,1}(-u)= -u(u-1)
\times {\rm id},\label{3Unitarity}\\
&&\mbox{ Crossing
relation}:\,R_{1,2}(u)=-\sigma^y_1R_{1,2}^{t_1}(-u-1)\sigma^y_1.
\label{3crosing-unitarity}
\end{eqnarray}

The monodromy matrix and the corresponding transfer matrix of the
periodic XXX spin-$\frac12$ chain are respectively defined as
\index{monodromy matrix}
\begin{eqnarray}
&&T_0(u)=R_{0,N}(u-\theta_N)\cdots
R_{0,1}(u-\theta_1)=\left(\begin{array}{cc} A(u)\;\; & B(u) \\ C(u)\;\; & D(u)
\end{array}
\right),\label{1Tmono}\\ &&t(u)=tr_0T_0(u)=A(u)+D(u),\label{1trr}
\end{eqnarray}
with $\{\theta_j|j=1,\cdots,N\}$ being some generic site-dependent
inhomogeneity parameters. With the Yang-Baxter equation we can show that $[t(u),t(v)]=0$. The Hamiltonian of
the XXX spin-$\frac12$ chain is thus expressed as\index{Heisenberg
model}\index{XXX spin chain}
\begin{eqnarray}
H=\frac12\sum_{j=1}^N\vec{\sigma}_j\cdot\vec{\sigma}_{j+1}=\left.\frac{\partial\ln
t(u)}{\partial u}\right|_{u=0,\{\theta_j=0\}}-\frac12N,\label{1h}
\end{eqnarray}
with the periodic boundary condition
$\vec{\sigma}_{N+1}\equiv\vec{\sigma}_{1}$.

In order to get some functional
relations of the transfer matrix, we evaluate the transfer matrix
$t(u)$ at the particular points $u=\theta_j$ and
$u=\theta_j-1$. Let us apply the initial condition of the $R$-matrix to
express the transfer matrix $t(\theta_j)$ as
\begin{eqnarray}
t(\theta_j)&=&tr_{0}\{R_{0,N}(\theta_j-\theta_N)\cdots
R_{0,j+1}(\theta_j-\theta_{j+1}) \nonumber
\\   &&\times P_{0,j} R_{0,j-1}(\theta_j-\theta_{j-1})\cdots
R_{0,1}(\theta_j-\theta_1)\}\nonumber\\[2pt]
&=&R_{j,j-1}(\theta_j-\theta_{j-1})\cdots R_{j,1}(\theta_j-\theta_1)
\nonumber
\\  &&\times tr_{0}\{R_{0,N}(\theta_j-\theta_N)\cdots
R_{0,j+1}(\theta_j-\theta_{j+1})P_{0,j}\}\nonumber\\
&=&R_{j,j-1}(\theta_j-\theta_{j-1})\cdots R_{j,1}(\theta_j-\theta_1)
\nonumber
\\  &&\times  R_{j,N}(\theta_j-\theta_N)\cdots R_{j,j+1}(\theta_j-\theta_{j+1}).
\end{eqnarray}
In deriving the above equation, the initial condition (\ref{3Int-R})
of the $R$-matrix plays a key role, which allows us to rewrite the
transfer matrix as a product of $R$-matrices at the special spectral
parameter points $\theta_j$. The transfer matrix $t(\theta_j)$ is a
reduced monodromy matrix if the $j$-th quantum space is treated as
the auxiliary space.

The
crossing relation (\ref{3crosing-unitarity}) makes it possible to express
the transfer matrix $t(\theta_j-1)$ as
\begin{eqnarray}
t(\theta_j-1)&=&tr_0\{ R_{0,N}(\theta_j-\theta_N-1)\cdots
R_{0,1}(\theta_j-\theta_1-1)\}\nonumber\\
&=&(-1)^N\,tr_0\{\sigma^y_0 R^{t_0}_{0,N}(-\theta_j+\theta_N)\cdots
R^{t_0}_{0,1}(-\theta_j+\theta_1)\sigma^y_0\}\nonumber\\
&=&(-1)^N\,tr_0\{ R_{0,1}(-\theta_j+\theta_1)\cdots
R_{0,N}(-\theta_j+\theta_N)\}\nonumber\\
&=&(-1)^{N}R_{j,j+1}(-\theta_j+\theta_{j+1})\cdots
R_{j,N}(-\theta_j+\theta_N)\nonumber\\ &&\times
R_{j,1}(-\theta_j+\theta_1)\cdots R_{j,j-1}(-\theta_j+\theta_{j-1}).
\end{eqnarray}

Using the unitary relation (\ref{3Unitarity}), we
have\index{operator product identities}
\begin{eqnarray}
&&t(\theta_j) t(\theta_j-1)=a(\theta_j)d(\theta_j-1),\quad
j=1,\cdots,N,\label{1operator-id-1}\\
&&a(u)=\prod_{j=1}^N{(u-\theta_j+1)},\quad
d(u)=\prod_{j=1}^N(u-\theta_j).
\end{eqnarray}
The homogeneous analogue of (\ref{1operator-id-1}) reads
\begin{eqnarray}
\frac{\partial^l}{\partial
u^l}\{t(u)t(u-1)-a(u)d(u-1)\}|_{u=0,\{\theta_j=0\}}=0,{~}
l=0,\cdots,N-1.
\end{eqnarray}

Applying (\ref{1operator-id-1}) to an eigenstate of $t(u)$, the
corresponding eigenvalue $\Lambda(u)$ thus
satisfies\index{functional relations}
\begin{eqnarray}
\Lambda(\theta_j)
\Lambda(\theta_j-1)=a(\theta_j)d(\theta_j-1),\quad
j=1,\cdots,N.\label{3Eigen-id-xxx}
\end{eqnarray}
In addition, from the definition of the transfer matrix (\ref{1trr}) it is easy to show that
\begin{eqnarray}
\Lambda(u) \mbox{ is a degree $N$ polynomial of $u$},\label{xxx-an}
\end{eqnarray}
with the asymptotic behavior\index{asymptotic behavior}
\begin{eqnarray}
\Lambda(u)=2u^N+\cdots.\label{xxx-asxxx}
\end{eqnarray}
The relations (\ref{3Eigen-id-xxx})-(\ref{xxx-asxxx}) determine the spectrum of the model completely.

We can easily demonstrate
that for any given parameter $\phi$, the following inhomogeneous
T-Q relation satisfies
(\ref{3Eigen-id-xxx})-(\ref{xxx-asxxx}) and therefore characterizes
the spectrum of the transfer matrix $t(u)$ of the periodic XXX
spin-$\frac12$ chain completely
\begin{eqnarray}
\hspace{-0.8truecm}\Lambda(u)&=&e^{i\phi}a(u)\frac{Q(u-1)}{Q(u)}+e^{-i\phi}d(u)\frac{Q(u+1)}{Q(u)}
+2(1-\cos\phi)\frac{a(u)d(u)}{Q(u)},\label{13qqq}\\
&&{~}Q(u)=\prod_{j=1}^N(u-\mu_j),\label{q}
\end{eqnarray}
provided that the Bethe roots $\{\mu_j|j=1,\cdots,N\}$ satisfy the
Bethe Ansatz equations (BAEs)
\begin{eqnarray}
e^{i\phi}a(\mu_j)Q(\mu_j-1)+e^{-i\phi}d(\mu_j)Q(\mu_j+1)=2(\cos\phi-1)a(\mu_j)d(\mu_j),\label{bae1xxxin}
\end{eqnarray}
and the selection rules $\mu_j\neq\mu_l$, $\mu_j\neq \theta_l,
\theta_l-1$.

\noindent{\bf Proposition 1:} {\it Each solution of
(\ref{3Eigen-id-xxx})-(\ref{xxx-asxxx}) can be parameterized in
terms of the inhomogeneous $T-Q$ relation (\ref{13qqq}) with a polynomial
$Q$-function (\ref{q}).}
\\[6pt]
 \noindent Proof: Given a degree $N$ polynomial
$\Lambda(u)$ satisfying (\ref{3Eigen-id-xxx})-(\ref{xxx-asxxx}), we
seek the degree $N$ polynomial solution of  $Q$-function satisfying
the equation
\begin{eqnarray}
Q(u)\Lambda(u)=e^{i\phi}a(u)Q(u-1)+e^{-i\phi}d(u)Q(u+1)+2(1-\cos\phi)a(u)d(u).\label{3qqq}
\end{eqnarray}
We note that the above equation is a polynomial one of degree $2N$. If
the equation holds at $2N+1$ independent points of $u$, the equation
must also hold for arbitrary $u$. Obviously, the above equation
holds for $u\to\infty$. In addition, as
$d(\theta_j)=a(\theta_j-1)=0$, we readily obtain that
\begin{eqnarray}
&&Q(\theta_j)\Lambda(\theta_j)=e^{i\phi}a(\theta_j)Q(\theta_j-1),\label{qxxx-1}\\
&&Q(\theta_j-1)\Lambda(\theta_j-1)=e^{-i\phi}d(\theta_j-1)Q(\theta_j).\label{qxxx-2}
\end{eqnarray}
From (\ref{3Eigen-id-xxx}) we deduce that only one of (\ref{qxxx-1}) and (\ref{qxxx-2}) is independent. Obviously, (\ref{qxxx-1}) (or equivalently (\ref{qxxx-2})) allows a degree $N$ polynomial solution of $Q(u)$
\begin{eqnarray}
Q(u)=u^N+\sum_{n=0}^{N-1}\tilde I_nu^n=\prod_{j=1}^N(u-\mu_j).\label{qxxxansatz}
\end{eqnarray}
Substituting the above Ansatz into (\ref{qxxx-1}) we have $N$ linear equations for the $N$ coefficients $\{\tilde I_n|n=0,\cdots, N-1\}$ which have a unique solution for a given $\Lambda(u)$.
Taking $u=\mu_j$ in (\ref{3qqq}), we readily have the BAEs (\ref{bae1xxxin}). $\square$

\noindent {\bf Proposition 2:} {\it The functional relations
(\ref{3Eigen-id-xxx})-(\ref{xxx-asxxx}) are the sufficient and
necessary conditions to completely characterize the spectrum of the
transfer matrix (\ref{1trr}) with the $R$-matrix
(\ref{20140317-r}).}
\\[6pt]
\noindent Proof: Let us introduce the rotated monodromy matrix
\begin{eqnarray}
T_\phi(u)&=&\left(\begin{array}{cc} A_\phi(u)\;\; & B_\phi(u) \\ C_\phi(u)\;\; & D_\phi(u)
\end{array}\right)\nonumber\\
&=&\left(\begin{array}{cc} \cos\frac\phi2\;\; & -\sin\frac\phi2 \\ \sin\frac\phi2\;\; & \cos\frac\phi2
\end{array}
\right)\left(\begin{array}{cc} A(u)\;\; & B(u) \\ C(u)\;\; & D(u)
\end{array}
\right)\left(\begin{array}{cc} \cos\frac\phi2\;\; & -\sin\frac\phi2 \\ \sin\frac\phi2\;\; & \cos\frac\phi2
\end{array}
\right).
\end{eqnarray}
For each solution $\Lambda(u)$ of the functional
relations (\ref{3Eigen-id-xxx})-(\ref{xxx-asxxx}), in terms of the inhomogeneous
T-Q relation (\ref{13qqq}), we can construct the following eigenstate of the
transfer matrix:
\begin{eqnarray}
|\mu_1,\cdots,\mu_N\rangle=\prod_{j=1}^NB_\phi(\mu_j)|0\rangle,
\end{eqnarray}
while $|0\rangle$ is the all spin-up state.
Therefore, each solution of the
functional relations (\ref{3Eigen-id-xxx})-(\ref{xxx-asxxx})
corresponds to an eigenvalue of the transfer matrix. $\square$

The corresponding eigenvalues of
the Hamiltonian (\ref{1h}) read
\begin{eqnarray}
E=\left.\frac{\partial\ln
\Lambda(u)}{\partial u}\right|_{u=0,\{\theta_j=0\}}-\frac12N.\label{1lambda}
\end{eqnarray}
The numerical solutions of the BAEs (\ref{bae1xxxin}) and the
corresponding eigenvalues of the Hamiltonian (\ref{1h}) for $N=3$
and $N=4$ with an arbitrarily chosen $\phi$ are shown in Table 1.1
and Table 1.2 respectively. Those numerical simulations imply that
the inhomogeneous T-Q relation (\ref{13qqq}) and the BAEs
(\ref{bae1xxxin}) indeed give the correct and complete spectrum of
the periodic XXX spin-$\frac12$ chain model.
\begin{table}
\caption{\label{1N3} The numerical solutions of the BAEs (\ref{bae1xxxin}) for $N=3$, $\phi=-0.69315i$ and $\{\theta_j=0\}$. The eigenvalues $E_n$ calculated from
(\ref{1lambda}) are the same as those from the exact
diagonalization of the Hamiltonian (\ref{1h}). The symbol $n$ denotes the
number of the energy levels and $d$ indicates the number of degeneracy.}
\begin{center}
\begin{tabular}{ccccc} \hline\hline
$ \qquad  \qquad \mu_1 \qquad \qquad $ & $  \qquad\qquad \mu_2\qquad  \qquad$ & $ \qquad \qquad \mu_3 \qquad $ &
$   \qquad E_n  \qquad  $  & $ \quad d \quad $ \\
\hline
$-2.97259+1.15909i$ & $-2.51751-1.42184i$ & $-0.50990+0.26274i$ & $-1.50000$  & $2$ \\
$-2.97259-1.15909i$ & $-2.51751+1.42184i$ & $-0.50990-0.26274i$ & $-1.50000$ & $2$ \\
$-2.88462+0.00000i$ & $-1.55769-2.56650i$ & $-1.55769+2.56650i$ & $1.50000$ & $4$\\
\hline\hline \end{tabular}
\end{center}
\end{table}
\begin{table}
\caption{\label{1543N2}The numerical solutions of the BAEs (\ref{bae1xxxin}) for $N=4$, $\phi=-0.69315i$ and $\{\theta_j=0\}$. The eigenvalues $E_n$ calculated from
(\ref{1lambda}) are the same as those from the exact
diagonalization of the Hamiltonian (\ref{1h}). The symbol $n$ denotes the
number of the energy levels and $d$ indicates the number of degeneracy.}
\begin{center}{\scriptsize
\begin{tabular}{ cccccc} \hline\hline
$ \qquad \mu_1 \qquad $ & $ \qquad \mu_2 \qquad $ & $ \qquad \mu_3 \qquad $ & $ \qquad \mu_4 \qquad $ &  $ \quad E_n \quad $ & $ \quad d \quad $ \\
\hline
$-3.46085-2.04638i$ & $-3.46085+2.04638i$ & $-0.53915-0.28370i$ & $-0.53915+0.28370i$ & $-4.00000$ & $1$\\
$-3.49754-0.00000i$ & $-2.00000+2.49853i$ & $-2.00000-2.49853i$ & $-0.50246-0.00000i$ & $-2.00000$ & $3$\\
$-3.41695-0.01463i$ & $-2.20702+2.20734i$ & $-1.88461-2.68745i$ & $-0.49142+0.49474i$ & $-0.00000$ & $3$\\
$-3.41695+0.01463i$ & $-2.20702-2.20734i$ & $-1.88461+2.68745i$ & $-0.49142-0.49474i$ & $-0.00000$ & $3$\\
$-3.38446-2.02080i$ & $-3.38446+2.02080i$ & $-1.11571+0.00000i$ & $-0.11537+0.00000i$ & $0.00000$ & $1$\\
$-3.07558+1.25638i$ & $-3.07558-1.25638i$ & $-0.92442+3.56865i$ & $-0.92442-3.56865i$ & $2.00000$ & $5$\\
\hline\hline \end{tabular}}
\end{center}
\end{table}

In conclusion, we showed that the spectrum of the periodic Heisenberg spin-chain model can also be characterized by an inhomogeneous T-Q relation. This conclusion can be generalized to other quantum integrable models. We remark that in the present case, the inhomogeneous T-Q relation can be reduced to Baxter's homogeneous T-Q relation by taking $\phi=0$, and the degree of the Q polynomial can be reduced to $M$ with $0\leq M\leq N$ by taking some of the Bethe roots to be infinity. However, for most of the quantum integrable models without $U(1)$ symmetry, the inhomogeneous term is indeed irreducible.

\end{document}